\input harvmac
\input epsf
\noblackbox
\def\npb#1#2#3{{\it Nucl.\ Phys.} {\bf B#1} (#2) #3}
\def\plb#1#2#3{{\it Phys.\ Lett.} {\bf B#1} (#2) #3}
\def\prl#1#2#3{{\it Phys.\ Rev.\ Lett.} {\bf #1} (#2) #3}
\def\prd#1#2#3{{\it Phys.\ Rev.} {\bf D#1} (#2) #3}
\def\ijmp#1#2#3{{\it Int.\ J. Mod.\ Phys.} {\bf A#1} (#2) #3}
\def\jhep#1#2#3{{\it JHEP\/} {\bf #1} (#2) #3}
\def\atmp#1#2#3{{\it Adv.\ Theor.\ Math.\ Phys.} {\bf #1} (#2) #3.}
\newcount\figno
\figno=0
\def\fig#1#2#3{
\par\begingroup\parindent=0pt\leftskip=1cm\rightskip=1cm\parindent=0pt
\baselineskip=11pt
\global\advance\figno by 1
\midinsert
\epsfxsize=#3
\centerline{\epsfbox{#2}}
\vskip 12pt
{\bf Fig.\ \the\figno: } #1\par
\endinsert\endgroup\par
}
\def\figlabel#1{\xdef#1{\the\figno}}
\def\encadremath#1{\vbox{\hrule\hbox{\vrule\kern8pt\vbox{\kern8pt
\hbox{$\displaystyle #1$}\kern8pt}
\kern8pt\vrule}\hrule}}

\def\frac#1#2{{#1 \over #2}}
\def\p{\partial}
\def\bar#1{\overline{#1}}
\def\tilde#1{\widetilde{#1}}

\def\CJ{{\cal J}}                   
\def\CM{{\cal M}}                   \def\CN{{\cal N}}
                   
\def\CR{{\cal R}}

\def\R{{\bf R}}                     \def\S{{\bf S}}
                     \def\Z{{\bf Z}}
\def\EbarE{E_8\times\bar E_8}
\Title{\vbox{\baselineskip12pt
\hbox{hep-th/0002073}
\hbox{CALT-68-2255}
\hbox{CITUSC/00-004}
\hbox{PRA-HEP 00/02}
\vskip-.3in}}
{\vbox{\centerline{Casimir Effect Between World-Branes}
\bigskip
\centerline{in Heterotic M-Theory}}}
\medskip
\centerline{Michal Fabinger$^1$ and Petr Ho\v rava$^{2}$}
\medskip\bigskip
\centerline{\it $^1$Institute of Theoretical Physics, Charles University, 
18000 Prague 8, Czech Republic}
\centerline{\tt fabinger@fzu.cz}
\medskip
\centerline{\it $^2$CIT-USC Center for Theoretical Physics}
\centerline{\it California Institute of Technology, Pasadena, CA 91125, USA}
\centerline{\tt horava@theory.caltech.edu}
\baselineskip18pt
\medskip\bigskip\medskip
\noindent\baselineskip16pt
We study a non-supersymmetric $E_8\times\bar E_8$ compactification of 
M-theory on $\S^1/\Z_2$, related to the supersymmetric $E_8\times E_8$ theory 
by a chirality flip at one of the boundaries.  This system represents an 
M-theory analog of the D-brane anti-D-brane systems of string theory.  
Alternatively, this compactification can be viewed as a model of 
supersymmetry breaking in the ``brane-world'' approach to phenomenology.  We 
calculate the Casimir energy of the system at large separations, and show that 
there is an attractive Casimir force between the $E_8$ and $\bar E_8$ 
boundary.  We predict that a tachyonic instability develops at separations of 
order the Planck scale, and discuss the possibility that the M-theory 
fivebrane might appear as a topological defect supported by the $E_8\times\bar 
E_8$ system.  Finally, we analyze the eventual fate of the configuration, 
in the semiclassical approximation at large separations:  the two ends of the 
world annihilate by nucleating wormholes between the two boundaries.  
\Date{February 2000}
\lref\hw{P. Ho\v rava and E. Witten, ``Heterotic and Type I String Dynamics 
from Eleven Dimensions,'' \npb{460}{1996}{506}, hep-th/9510209.}
\lref\hweff{P. Ho\v rava and E. Witten, ``Eleven-Dimensional Supergravity on 
a Manifold with Boundary,'' \npb{475}{1996}{94}, hep-th/9603142.}
\lref\casimir{H.B.G. Casimir, ``On the Attraction Between Two Perfectly 
Conducting Plates,'' {\it Proc.\ Kon.\ Nederl.\ Akad.\ Wet.} {\bf 51} (1948) 
793.}
\lref\ewcy{E. Witten, ``Strong Coupling Expansion of Calabi-Yau 
Compactification,'' \npb{471}{1996}{135}, hep-th/9602070.}
\lref\phg{P. Ho\v rava, ``Gluino Condensation in Strongly Coupled Heterotic 
String Theory,'' \prd{54}{1996}{7561}, hep-th/9608019.}
\lref\rohm{R. Rohm, ``Spontaneous Supersymmetry Breaking in Supersymmetric 
String Theories,'' \npb{237}{1984}{553}.}
\lref\tomlenny{T. Banks and L. Susskind, ``Brane -- Anti-Brane Forces,'' 
hep-th/9511194.}
\lref\gsw{M.B. Green, J.H. Schwarz and E. Witten, {\it Superstring Theory} 
(Cambridge U. Press, 1987), Vol.~2, Appendix~8.A.}
\lref\runaway{M. Dine and N. Seiberg, ``Is the Superstring Weakly Coupled?,'' 
\plb{162}{1985}{299}, ``Is the Superstring Semiclassical?,'' in: 
{\it Unified String Theories}, eds: M.B. Green and D.J. Gross 
(World Scientific, 1986).}
\lref\bdstrong{T. Banks and M. Dine, ``Couplings and Scales in Strongly 
Coupled Heterotic String Theory,'' \npb{479}{1996}{173}, hep-th/9605136.}
\lref\phbh{P. Ho\v rava, ``Two-Dimensional Stringy Black Holes with One 
Asymptotically Flat Domain,'' \plb{289}{1992}{293}, hep-th/9203031.}
\lref\hhk{J.A. Harvey, P. Ho\v rava and P. Kraus, ``D-Sphalerons and the 
Topology of String Configuration Space,'' hep-th/0001143.}
\lref\coleman{S. Coleman, ``The Uses of Instantons,'' in: {\it Aspects of 
Symmetry} (Cambridge U. Press, 1985).}
\lref\hhkfalse{J.A. Harvey, P. Ho\v rava and P. Kraus, to appear.}
\lref\ewkk{E. Witten, ``Instability of the Kaluza-Klein Vacuum,'' 
\npb{195}{1982}{481}.}
\lref\ewk{E. Witten, ``D-Branes and K-Theory,'' \jhep{9812}{1998}{019}, 
hep-th/9810188.}
\lref\phk{P. Ho\v rava, ``Type IIA D-Branes, K-Theory, and Matrix Theory,'' 
\atmp{2}{1998}{1373}, hep-th/9812135.}
\lref\sen{A. Sen, ``Non-BPS States and Branes in String Theory,'' 
hep-th/9904207, and references therein.}
\lref\mm{R. Minasian and G. Moore, ``K-Theory and Ramond-Ramond Charge,'' 
\jhep{9711}{1997}{002}, hep-th/9710230.}
\lref\wm{E. Witten and G. Moore, ``Self-Duality, Ramond-Ramond Fields, and 
K-Theory,'' hep-th/9912279.}
\lref\sens{A. Sen, ``Universality of the Tachyon Potential,'' 
\jhep{9912}{1999}{027}, hep-th/9911116; A. Sen and B. Zwiebach, ``Tachyon 
Condensation in String Field Theory,'' hep-th/9912249.}
\lref\yi{P. Yi, ``Membranes from Five-Branes and Fundamental Strings from 
D$p$ Branes,'' \npb{550}{1999}{214}, hep-th/9901159.}
\lref\mtw{C.W. Misner, K.S. Thorne and J.A. Wheeler, {\it Gravitation} 
(Freeman, New York, 1973).}
\lref\flurry{A. Lukas, B.A. Ovrut, K.S. Stelle and D. Waldram, ``The Universe 
as a Domain Wall,'' \prd{59}{1999}{086001}, hep-th/9803235; 
N. Arkani-Hamed, S. Dimopoulos and G. Dvali, ``The Hierarchy Problem and New 
Dimensions at a Millimeter,'' \plb{429}{1998}{263}, hep-ph/9803315; 
I. Antoniadis, N. Arkani-Hamed, S. Dimopoulos and G. Dvali, ``New Dimensions 
at a Millimeter and Superstrings at a TeV,'' \plb{436}{1998}{257}, 
hep-ph/9804398; 
Z. Kakushadze and S.-H.H. Tye, ``Brane World,'' \npb{548}{1999}{180}, 
hep-th/9809147; 
L. Randall and R. Sundrum, ``A Large Mass Hierarchy from a Small Extra 
Dimension,'' \prl{83}{1999}{3370}, hep-ph/9905221, to name a few.}
\lref\klt{H. Kawai, D.C. Lewellen and S.-H.H. Tye, ``Classification of 
Closed Fermionic String Models,'' \prd{34}{1986}{3794}.}
\lref\dixh{L.J. Dixon and J.A. Harvey, ``String Theories in Ten Dimensions 
Without Spacetime Supersymmetry,'' \npb{274}{1986}{93}.}
\lref\ewtools{E. Witten, ``Topological Tools in Ten-Dimensional Physics,'' 
\ijmp{1}{1986}{39}, also in: {\it Unified String Theories}, eds: M.B. Green 
and D.J. Gross (World Scientific, 1986).}
\lref\manyf{N. Arkani-Hamed, S. Dimopoulos, G. Dvali and N. Kaloper, 
``Manyfold Universe,'' hep-th/9911386.}
\newsec{Introduction}

M-theory on an eleven-dimensional manifold $\CM$ with non-empty boundary 
$\p\CM$ is described at long distances by bulk supergravity coupled to super 
Yang-Mills theory on $\p\CM$ \refs{\hw,\hweff}.  The choice of the gauge group 
in the boundary sector is determined by an anomaly cancellation argument: each 
boundary component supports one copy of the $E_8$ supermultiplet.  Thus, for 
example, the two boundary components of the supersymmetric $\Z_2$ orbifold 
$\R^{10}\times\S^1/\Z_2$ support one copy of $E_8$ each, and this orbifold 
describes the strongly coupled regime of the $E_8\times E_8$ heterotic string 
on $\R^{10}$ \hw . 

Even though the anomaly cancellation mechanism of \refs{\hw,\hweff} uniquely 
determines the Yang-Mills gauge group at each boundary component to be $E_8$, 
in order to fully specify the boundary theory we still have a discrete choice 
to make.  The ten-dimensional Yang-Mills supermultiplet contains a 
Majorana-Weyl gaugino $\chi$, which satisfies one of two possible chirality 
conditions, 
\eqn\eechirality{\chi=\pm\Gamma_{11}\chi.}
Once a choice of the sign in \eechirality\ is made, the chirality of the 
boundary conditions on the bulk gravitino is also uniquely determined.  

Since the anomaly cancellation argument works locally near each component of 
the boundary, the discrete choice of chirality in \eechirality\ can be made 
independently at each boundary component.  On a manifold $\CM$ with two 
boundary components, we thus have two distinct options: $(+,+)$ and $(+,-)$, 
depending on whether the two chiralities agree or disagree.  

Consider again $\CM=\R^{10}\times\S^1/\Z_2$ with a flat, direct-product 
metric.  In the case of the $(+,+)$ boundary conditions, the two boundaries 
break the same half of the original supersymmetry, and we obtain the strongly 
coupled limit of $E_8\times E_8$ heterotic string theory presented in \hw .  
In the $(+,-)$ case, each boundary component breaks a separate set of sixteen 
supercharges, leading to a configuration with gauge symmetry $E_8\times E_8$ 
but no supersymmetry.  We will refer to the $(+,-)$ case as the ``$\EbarE$ 
compactification,'' in order to indicate the opposite choice of chirality in 
the second $E_8$ factor, and to avoid any possible confusion with the 
supersymmetric $E_8\times E_8$ compactification of \hw .  It is this 
non-supersymmetric $\EbarE$ theory that will be the subject of the present 
paper.  

Since supersymmetry is completely broken in the $\EbarE$ model, the distance 
$L$ between the two boundaries is no longer an exact modulus, and the theory 
develops a non-trivial potential for $L$.  (Furthermore, the flat metric on 
$\CM$ will also be modified by quantum corrections.)  On these grounds, one 
can expect an attractive or repulsive force between the two boundaries that 
are initially at some separation $L$.  We will analyze the force between the 
boundaries in the long-wavelength approximation, at separations much larger 
than the eleven-dimensional Planck length, $L\gg \ell_{11}$.  

In the course of this paper, we will keep in mind two possible applications 
of the $\EbarE$ system.  

First of all, we observe that the $\EbarE$ system can be thought of as an 
analog of the unstable D$p$-D$\bar p$ brane systems recently much studied in 
string theory \refs{\sen,\ewk,\phk}.  A system of D$p$-D$\bar p$ brane pairs 
is unstable, and tends to annihilate to the vacuum.  Indeed, the system 
develops an open-string tachyon at D$p$-D$\bar p$ separations smaller 
than the string scale.  This tachyon behaves as a Higgs field, and the Higgs 
mechanism corresponds to the world-volume description of the brane-antibrane 
annihilation.  In the process of its annihilation, the unstable system can 
leave behind a bound state in the form of a lower-dimensional stable D-brane 
that appears as a defect on the worldvolume of the original unstable system.  
All stable D-branes can be described in this way as topological defects in 
a universal unstable system of spacetime-filling branes \refs{\ewk,\phk}.  
Underlying this construction is a deep relation between D-brane charges, 
RR fields, and K-theory \refs{\mm,\ewk,\phk,\wm}.  As one of the points of 
this paper, we will try to convince the reader that the $E_8\times\bar E_8$ 
system is indeed a rather close M-theoretic analog of such unstable 
D$p$-D$\bar p$ systems of Type~II and Type~I string theory, and in fact 
exhibits some properties expected of the universal unstable system in 
M-theory.  

Alternatively, one can compactify the $\EbarE$ model on $\R^4\times
\S^1/\Z_2\times Y$,%
\foot{Here $Y$ could be a Calabi-Yau manifold with a characteristic scale 
much smaller than the size of the $\S^1/\Z_2$.}
and think of one of the $E_8$ boundaries as a brane-world on the boundary of 
an effectively five-dimensional spacetime.  In fact, it was this 
compactification of the supersymmetric $(+,+)$ model that was the direct 
predecessor \refs{\ewcy,\bdstrong,\phg} of the brane-world scenarios with 
large extra dimensions, and stimulated much of the recent flurry of interest 
in that area \flurry .  Similarly, the $\EbarE$ model provides an intriguing 
example of supersymmetry breaking in the brane-world scenario, in a context 
fully embedded into M-theory.  One could use the $\EbarE$ model to address 
some of the important issues expected to arise in the brane-world physics, 
such as the dilaton runaway problem \runaway\ (or its M-theoretic dual, 
``radius runaway'' problem \refs{\bdstrong,\phg}), radius stabilization, and 
the scale of supersymmetry breaking.  In addition, our analysis of the 
$\EbarE$ model will allow us to raise some important new issues -- most 
notably, the issue of a catastrophic instability of some brane-world 
compactifications due to false vacuum decay.  

\newsec{Casimir Effect Between Two Ends of the World in M-Theory}

\subsec{The $\EbarE$ model}

Consider M-theory in $\R^{11}$ in a coordinate system $x^M$, $M=0,\ldots 
,10$, with a flat metric $g_{MN}=\eta_{MN}\equiv{\rm diag}\,(-+\cdots+)$, and 
with a boundary along $x^{10}=0$.  It is convenient to think of this model as 
a $\Z_2$ orbifold of M-theory in $\R^{11}$, where the orbifold group acts by 
$x^{10}\rightarrow -x^{10}$.  In this picture, the boundary conditions on the 
gravitino are induced from the orbifold condition 
\eqn\eebcgrino{\psi(-x^{10})=\Gamma_{10}\psi(x^{10}).}
(Here we are using a condensed notation, $\psi^\alpha=\psi^\alpha_\mu dx^\mu$ 
for the gravitino, with $\alpha$ being the 32-component Majorana spin index.)  
The boundary condition \eebcgrino\ breaks one half of the original 
supersymmetry, and defines what we mean by the ``+ chirality.''  The boundary 
supports a Yang-Mills supermultiplet $(A_A^a,\chi^a)$, where $x^A, A=0,\ldots 
9$ are the coordinates along the boundary, $a$ denotes the adjoint 
representation of $E_8$, and the gaugino $\chi^a$ satisfies $\chi^a=
\Gamma_{11}\chi^a$, with the role of $\Gamma_{11}$ played by $\Gamma_{10}$.  

Imagine bringing in another boundary component adiabatically from infinity to 
a finite distance $x^{10}=L$, with the opposite choice of boundary 
conditions.  (This corresponds to the $(+,-)$ model of the introduction.)  It 
is again useful to think of this compactification as a $\Z_2$ orbifold of 
M-theory compactified to ten dimensions on $\tilde\CM=\R^{10}\times\S^1$ with 
radius
\eqn\eeradius{R_{10}=L/\pi,}
and with the gravitino boundary condition at $x^{10}=L$ induced from the 
orbifold condition 
\eqn\eebcgrinotwo{\psi(L-x^{10})=-\Gamma_{10}\psi(L+x^{10}).}
Combining \eebcgrino\ and \eebcgrinotwo , the gravitino is found to be 
antiperiodic around the $\S^1$ factor of $\tilde\CM$, 
\eqn\eeantipgrino{\psi(x^{10}+2\pi R_{10})=-\psi(x^{10}),}
and our model can be formally thought of as a $\Z_2$ orbifold of M-theory on 
$\tilde\CM$ with this non-supersymmetric choice of the spin structure.%
\foot{In general, we do not understand M-theory well enough to be able 
to determine how its non-supersymmetric orbifolds should be constructed.   
However, in the case of our interest, each boundary component separately 
breaks only a half of the original supersymmetry.  A mild assumption of 
cluster decomposition is sufficient to determine what happens at each 
boundary, as long as their separation is large.}
(Compactifications of string theory on $\S^1$ with the non-supersymmetric 
spin structure were first studied by Rohm \rohm .)  

\fig{Two compactifications of M-theory on $\S^1/\Z_2$: (a) The supersymmetric 
$E_8\times E_8$ model, and (b) the nonsupersymmetric $\EbarE$ model.  The 
arrows along the spacetime boundaries schematically denote the chiralities of 
the boundary $E_8$ Yang-Mills supermultiplets.}{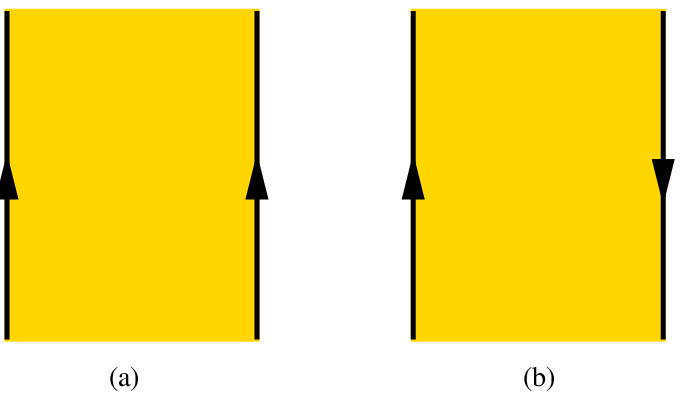}{3truein}
Each boundary component separately supports a copy of the $E_8$ Yang-Mills 
supermultiplet, and breaks one half of the original supersymmetry.  The 
low-energy Lagrangian of the system at large $L$ is that of eleven-dimensional 
bulk supergravity coupled to one $E_8$ multiplet at each boundary component, 
and can in principle be constructed systematically as an expansion in the 
powers of the eleven-dimensional Planck length $\ell_{11}$ (or, more 
precisely, as a long-wavelength expansion in the powers of the dimensionless 
parameter $\ell_{11}/L$), much like in \hweff .  

\subsec{Casimir force between the boundaries}

Since the model breaks all supersymmetry, the size $L$ of $\S^1/\Z_2$ is 
not a modulus, and quantum effects will generate a non-trivial potential for 
$L$.  This potential leads to a force between the two boundaries, which can be 
either repulsive or attractive.  In other words, the nonsupersymmetric 
$\EbarE$ system will exhibit an M-theoretical analog of the Casimir effect 
\casimir.  In this section we will determine the leading behavior of the 
Casimir force at large separations $L$ between the boundaries. 

If the force turns out to be repulsive, the eventual fate of the system will 
be uninteresting:  the system will decompactify and sixteen supersymmetries 
will be restored as $L\rightarrow\infty$.  In contrast, the case of an 
attractive force would be much more interesting.  In that case, one could 
imagine setting up adiabatically an initial configuration with a very large 
separation between the two boundaries, and then letting the system evolve.  
The two boundaries will start attracting each other, and will presumably soon 
reach the regime of $L$ of order the Planck length $\ell_{11}$ where our 
supergravity approximation is no longer valid.  Still, the question of the 
final fate of the system makes perfect sense, and should have a well-defined 
answer in the full quantum M-theory despite our current inability to determine 
it due to our limited understanding of M-theory in the strongly coupled 
regime.  Alternatively, one might hope that the potential has a minimum 
at some value of $L$ that is large enough so that perturbation theory can 
still be used to analyze the resulting vacuum; this option would certainly 
be interesting phenomenologically.  

We will now demonstrate that the Casimir force at large separation $L$ is 
indeed attractive.  Our calculation will proceed as follows.  We start with 
the $\EbarE$ model on $\CM=\R^{10}\times\S^1/\Z_2$ with a flat, direct-product 
metric
\eqn\eeclassical{ds_0^2=\eta_{AB}dx^Adx^B+L^2dz^2,}
where $x^A$, $A=0,\ldots 9$ are coordinates on $\R^{10}$, and we have 
introduced a rescaled coordinate $z$ on the $\S^1/\Z_2$ factor such that 
$z\in[0,1]$.  We assume that the distance $L$ between the boundaries is 
constant and large in Planck units.  The geometry \eeclassical\ represents a 
classical solution of the theory.  Quantum fluctuations of the fields on $\CM$ 
generate a non-zero expectation value $\langle T_{MN}\rangle$ of the 
energy-momentum tensor, which then modifies the classical flat static geometry 
of $\CM$.  At large separations $L$, this effect can be systematically studied 
in the long-wavelength expansion, i.e., in the perturbation theory in powers 
of $\ell_{11}/L$.  In this paper, we will only be interested in the 
leading-order perturbative correction to the flat geometry of $\CM$.  It is 
easy to show that the first non-zero contribution to $\langle T_{MN}\rangle$ 
will come from one loop in the supergravity sector, due to the mismatch in the 
boundary conditions for bosons and fermions in the supergravity multiplet.%
\foot{The boundary Yang-Mills multiplets only contribute to 
$\langle T_{MN}\rangle$ at higher orders in the long-wavelength expansion, 
and will not enter our calculation.}
Thus, our aim is to first calculate $\langle T_{MN}\rangle$ at one loop, and 
then determine the response of the metric on $\R^{10}\times\S^1/\Z_2$ 
to the leading non-trivial order in our long-wavelength expansion.  

Before actually calculating the first quantum correction 
$\langle T_{MN}\rangle$ to the vanishing energy momentum tensor of 
\eeclassical , notice first that its possible form is severely constrained.  
First of all, the Poincar\'e symmetry of the background metric \eeclassical\ 
implies that $\langle T_{MN}\rangle$ takes the form
\eqn\eetformone{\langle T_{MN}\rangle dx^Mdx^N=-E(z)\eta_{AB}dx^Adx^B+F(z)L^2
dz^2,}
with $E(z)$ and $F(z)$ are in general some functions of $z$.  Furthermore, the 
condition of energy-momentum conservation implies that $F$ is a constant 
independent of $z$, but does not restrict the functional dependence of $E$ on 
$z$.  In order to determine $E(z)$, notice that in our system, the one-loop 
energy-momentum tensor in the flat background \eeclassical\ has to be 
traceless.  This implies that $F=10E(z)$, and therefore $E(z)=E_0$ is a 
constant and the energy-momentum tensor \eetformone\ takes the following 
general form, 
\eqn\eetform{\langle T_{MN}\rangle dx^Mdx^N=-E_0\,(\eta_{AB}dx^Adx^B-10L^2
dz^2).}
The remaining constant $E_0$ plays the role of the vacuum energy density in 
the eleven-dimensional theory, and can be efficiently determined by 
Kaluza-Klein reducing the theory from $\R^{10}\times\S^1/\Z_2$ to $\R^{10}$, 
and calculating the effective one-loop energy-momentum tensor 
$\langle T_{AB}\rangle_{10}$ of all the KK modes in $\R^{10}$.  By Poincar\'e 
symmetry, we have 
\eqn\eepoincb{\langle T_{AB}\rangle_{10}=-\tilde E_0\,\eta_{AB},}
where $\tilde E_0$ is the the vacuum energy density in ten dimensions, or 
the one-loop effective cosmological constant.  $\tilde E_0$ is related to the 
vacuum energy density  $E_0$ in eleven dimensions by 
\eqn\eeenergies{\tilde E_0=L\int dz\,E_0=LE_0.}
The one-loop energy density $\tilde E_0$ is conveniently given by 
\eqn\eetilde{\tilde E_0=-\int\frac{d^{10}p}{(2\pi)^{10}}\sum_{p_i}(-1)^{F_i}
\int_0^\infty\frac{d\ell}{2\ell}e^{-(p^2+p_i^2)\ell/2},}
where the sum over $p_i$ represents the sum over all Kaluza-Klein momenta 
as well as all possible polarizations in the supergravity multiplet, and 
${F_i}$ is the fermion number.  No UV regularization at $\ell\rightarrow 
0$ is needed as \eetilde\ will turn out to be finite.  From the 
ten-dimensional perspective, the KK reduction gives 128 bosonic polarizations 
at each mass level $\pi m/L$ for $m$ a positive integer, and 128 
fermionic polarizations at each mass level $\pi r/L$ for $r$ a positive 
odd-half-integer.  (Recall the antiperiodicity conditions on the fermions, 
\eeantipgrino.)  In addition, 64 out of the original 128 massless bosons 
also survive the orbifold projection from $\S^1$ to $\S^1/\Z_2$.  Altogether, 
\eetilde\ becomes 
\eqn\eetildei{\eqalign{\tilde E_0&=-64\int_0^\infty
\frac{d\ell}{2\ell}\frac{1}{(2\pi\ell)^5}\left(\sum_{m\in\Z}e^{-m^2\pi^2
\ell/2L^2}-\sum_{r\in\Z+\frac{1}{2}}e^{-r^2\pi^2\ell/2L^2}\right)\cr
&\quad{}=-64\int_0^\infty\frac{d\ell}{2\ell}\frac{1}{(2\pi\ell)^5}
\sum_{s\in\Z}(-1)^se^{-s^2\pi^2\ell/8L^2}\cr
&\qquad{}=-64\int_0^\infty\frac{d\ell}{2\ell}\frac{1}{(2\pi\ell)^5}
\theta_4(0|i\pi\ell/8L^2),\cr}}
where $\theta_4(u|t)$ is one of the Jacobi theta functions (our conventions 
for Jacobi theta functions are as in \gsw ).  Rescaling the loop parameter 
$\ell\rightarrow\tau$ such that all the dependence on $L$ is outside the 
integral, we thus obtain the following expression for the vacuum energy 
density per unit area of the boundary, 
\eqn\eecasimir{\tilde E_0=-\CJ\cdot\frac{1}{L^{10}},}
with the $L$-independent factor $\CJ$ given by the integral 
\eqn\eeintegral{\CJ=\frac{1}{2^{15}}\int_0^\infty\frac{d\tau}{\tau^6}
\theta_4(0|i\tau).}

It is easy to demonstrate that $\CJ$ is convergent and positive.  First, 
change the variables to $t=1/\tau$, and use the modular properties of 
the Jacobi theta functions, $\theta_4(0|T)=(-iT)^{-1/2}\theta_2(0|-1/T)$ to 
obtain
\eqn\eeintegtwo{\CJ=\frac{1}{2^{15}}\int_0^\infty dt\,t^{9/2}\theta_2(0|it).}
The theta function $\theta_2(0|it)$ is positive definite for real $t$, and 
decays exponentially as $t\rightarrow\infty$.  Therefore, the integral 
over $\tau$ in \eeintegral\ is convergent and positive.  This shows that 
the vacuum energy density $\tilde E_0$ per unit boundary area as given by 
\eecasimir\ is negative.  

Thus, we have demonstrated that the Casimir effect between the boundaries of 
the $\EbarE$ model induces, in the leading order of the long-wavelength 
approximation, a negative cosmological constant.  It is tempting to conclude 
that the negative ten-dimensional cosmological constant implies an attractive 
force between the two boundaries.  Although this conclusion will turn out to 
be correct in our case (as we will see in detail in section~2.3), it 
cannot be reached with the mere knowledge of $\tilde E_0$ and requires a more 
detailed information about the energy-momentum tensor in eleven dimensions.  
Indeed, it is not the sign of the vacuum energy density, but rather the sign 
of $T_{zz}$ that determines whether the force between the boundaries is 
attractive or repulsive.  Using \eetform , \eecasimir , and \eeenergies, we 
obtain the one-loop energy-momentum tensor in eleven dimensions,%
\foot{This expression for the energy-momentum tensor can also be obtained by 
a direct one-loop calculation of the expectation value of the composite 
operator $T_{MN}$ in eleven dimensions.  This calculation reproduces our 
result (2.15), and we leave it as an exercise for the reader.}
\eqn\eecaseleven{\langle T_{MN}\rangle dx^Mdx^N=\frac{\CJ}{L^{11}}(\eta_{AB}
dx^Adx^B-10L^2dz^2).}
The Casimir force $\CF$ between the boundaries (per unit boundary area) is 
given by 
\eqn\eezeezee{\CF=\langle T_{\hat z\hat z}\rangle=-\frac{10\CJ}{L^{11}}\ \ <0,}
where $T_{\hat z\hat z}$ is the $zz$ component of the energy-momentum tensor 
\eecaseleven\ in the orthonormal vielbein.  It is reassuring that in our model 
the Casimir force $\CF$ can also be obtained from the response of the energy 
density per unit boundary area to changing $L$, 
\eqn\eecasforce{\CF=-\frac{\p\tilde E_0}{\p L}=-\frac{10\CJ}{L^{11}}.}
We conclude that the leading-order Casimir force exerted on the boundaries in 
the $\EbarE$ model at large $L$ is indeed attractive.  Notice that this force 
exhibits the typical Casimir-like scaling (as $L^{-D}$ in $D$ spacetime 
dimensions) familiar from the conventional Casimir effect in electrodynamics 
\casimir .  

\subsec{Backreaction from the geometry}

Imagine an initial configuration $\R^{9}\times\S^1/\Z_2$ with the two 
boundaries at some large constant initial separation $L_0$, set up by 
starting in flat $\R^{10}$ and adiabatically bringing the boundaries in from 
infinity.  The attractive Casimir force whose existence was demonstrated in 
section~2.2 suggests that as this initial configuration evolves with time, 
the boundaries should start moving closer together towards smaller values of 
$L$.  This is similar to the case of a D$p$-D$\bar p$ pair in string theory, 
but there are also some marked differences.  Unlike the case of a D$p$-brane, 
the effective theory on the $E_8$ boundary in M-theory does not contain a 
scalar that would describe the transverse movement of the boundary.  Hence, 
if the two boundaries are to move closer together under the influence of the 
Casimir force, it has to be due to a backreaction of the bulk metric to the 
non-zero Casimir energy-momentum tensor induced by the boundaries.  

We will now analyze this response of the metric to the non-zero $\langle 
T_{MN}\rangle$ of \eecaseleven , in the leading order in the long-wavelength 
expansion.  Consider the following general form of the metric on 
$\R\times\R^9\times\S^1/\Z_2$, 
\eqn\eemetansatz{ds^2=-dt^2+a^2(t)g_{ij}dx^idx^j+L^2(t)dz^2,}
where we have again used the rescaled coordinate $z$ along $\S^1/\Z_2$, with 
$z\in[0,1]$.  The indices $i,j=1,\ldots 9$ parametrize the spacelike slice 
(topologically $\R^9$) of the boundary geometry.  The metric $g_{ij}$ on 
$\R^9$ is constrained by the symmetries of the problem to be of constrant 
curvature, i.e., its Ricci tensor $\tilde R_{ij}$ satisfies 
$\tilde R_{ij}=kg_{ij}$.  The initial configuration at $t=0$ corresponds to 
\eqn\eeinitial{ds^2_{\R^9\times\S^1/\Z_2}=g_{ij}dx^idx^j+L_0^2dz^2,}
and we will study its response to the Casimir energy-momentum tensor at 
small $t>0$, in the leading order in the eleven-dimensional Newton constant 
$G_{11}\sim\ell_{11}^9$.  In the metric \eeinitial\ we had to allow for the 
possibility that the metric on $\R^9$ is not flat; in fact, as we will see 
below, its constant curvature $k$ turns out to be non-zero at order $G_{11}$.  

At zeroth-order, the metric is flat and the three-form gauge field $C$ is 
zero, and we do not have to worry about corrections to Einstein's 
equations from higher-power curvature terms or the $C$-dependent terms in the 
Lagrangian.  Thus, the equations of motion at first order in $G_{11}$ are 
simply 
\eqn\eeeom{R_{MN}=8\pi G_{11}\langle T_{MN}\rangle.}
Given our one-loop result for the energy-momentum tensor \eecaseleven , we 
take $\langle\ T_{MN}\rangle$ in the form
\eqn\eetmnansatz{\langle T_{MN}\rangle dx^Mdx^N=\frac{\CJ}{L^{11}(t)}(-dt^2
+a^2(t)g_{ij}dx^idx^j-10L^2(t)dz^2),}
where $L$ is now allowed to depend on $t$.  Notice that this adiabatic 
assumption is compatible with the requirement of energy-momentum conservation: 
the $T_{MN}$ of \eetmnansatz\ is conserved in the metric given by 
\eemetansatz .  The equations of motion \eeeom\ for \eemetansatz\ and 
\eetmnansatz\ lead to 
\eqn\eeeinst{\eqalign{-\frac{9\ddot a}{a}-\frac{\ddot L}{L}&=-8\pi G_{11}
\frac{\CJ}{L^{11}},\cr
8(\dot a)^2+a\ddot a+\frac{a}{L}\dot a\dot L+k&=8\pi G_{11}
\frac{a^2\CJ}{L^{11}},\cr
L\ddot L+\frac{L}{a}\dot a\dot L&=-80\pi G_{11}\frac{\CJ}{L^9}.\cr}}
Since we are looking for the leading backreaction  of the initial 
configuration \eeinitial\ to the $\langle T_{MN}\rangle$ given by 
\eetmnansatz\ at small $t>0$, we expand
\eqn\eemetexpand{\eqalign{L(t)=&L_0+\frac{1}{2}L_2t^2+\ldots,\cr
a(t)=&1+\frac{1}{2}a_2t^2+\ldots.\cr}}
Plugging this expansion into \eeeinst\ determines
\eqn\eebackreaction{\eqalign{k&=-\frac{16\pi\CJ G_{11}}{9L_0^{11}},\cr
L_2&=-\frac{80\pi\CJ G_{11}}{L_0^{10}},\cr
a_2&=\frac{88\pi\CJ G_{11}}{9L_0^{11}}.\cr}}
Thus, we reach the following conclusions:

(1)  At leading order in $G_{11}$, the spacetime geometry responds to the 
Casimir force by moving the boundaries closer together, i.e., $L(t)<L_0$ for 
(small) times $t>0$.  At the same time, the metric on the transverse $\R^9$ 
is rescaled by an increasing conformal factor $a(t)>1$.  

(2)  Interestingly, the naive initial configuration with $k=0$, corresponding 
to two flat boundaries at finite distance apart, is incompatible with the 
constraint part of Einstein's equations.  As we adiabatically bring in the 
second boundary from infinity, the geometry of the transverse $\R^9$ responds 
by curving with a constant negative curvature given by $k$ in 
\eebackreaction . 

\fig{The Casimir effect in the $\EbarE$ model.  According to \eemetexpand\ and 
\eebackreaction , the initial geometry on $\R^9\times\S^1/\Z_2$ with large 
initial separation $L_0$ evolves towards smaller $L$, while the boundary 
metric is getting rescaled.}{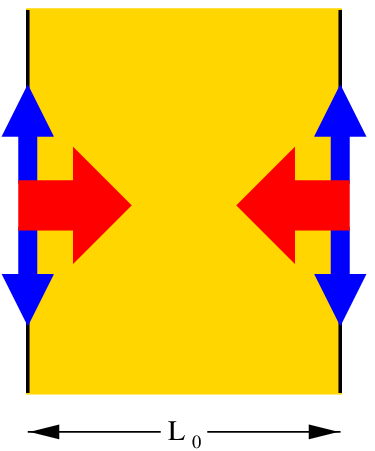}{2truein}

\subsec{Casimir effect on the open membrane}

The supersymmetric $E_8\times E_8$ compactification of M-theory on 
$\R^{10}\times\S^1/\Z_2$ describes the strongly coupled heterotic string 
theory in $\R^{10}$.  The heterotic string itself corresponds to the open 
membrane stretching between the two $E_8$ boundaries.  In this section we 
will study the open stretched membrane in the non-supersymmetric 
$\EbarE$ model, and will find close parallels with the spacetime picture of 
the Casimir effect.    

Consider an open membrane stretched between the two boundaries of spacetime, 
with worldvolume $\Sigma=\R^2\times\S^1/\Z_2$ parametrized by 
$(\sigma^m,\rho)$, $m=0,1$, and with $\rho\in[0,L]$.  In addition to 
$x^M(\sigma^m,\rho)$, the bulk worldvolume theory contains the spacetime 
spinor $\theta^\alpha(\sigma^m,\rho)$.  All boundary conditions are induced 
from the $\Z_2$ orbifold action on $x^M$, $\theta^\alpha$, and $\Sigma$.  In 
particular, the fermions satisfy 
\eqn\eebcmemf{\theta^\alpha(\sigma^m,-\rho)=\pm\Gamma_{10}\theta^\alpha(
\sigma^m,\rho),}
and similarly on the other boundary at $\rho=L$.    
This boundary condition \eebcmemf\ requires a sign choice, precisely 
correlated with the spacetime chirality choice \eechirality .  At each 
boundary, the bulk fields $x^M$ and $\theta^\alpha$ couple to a copy of the 
chiral $E_8$ current algebra at level one, whose chirality is uniquely 
determined by the choice of chirality in \eebcmemf .  Each boundary breaks one 
half of the original spacetime supersymmetry.  

In the $(+,+)$ model, both boundaries break the same half of the original 
supersymmetry.  The chiralities of the two $E_8$ current algebras agree, 
thus reproducing the characteristic chiral pattern of the heterotic string.  

In our non-supersymmetric $\EbarE$ model, corresponding to the $(+,-)$ 
chirality choice, the chiralities of the $E_8$ current algebras disagree, 
and each boundary breaks a separate half of the original supersymmetry.  
Due to this mismatch in the boundary conditions, we expect a worldvolume 
analog of the spacetime Casimir effect in the $\EbarE$ model.   Consider an 
open membrane with worldvolume $\R^2\times\S^1/\Z_2$ stretching along 
$x^1,\ldots x^8=0$ between the two boundaries.  We will calculate the leading 
correction $\tau$ to the membrane tension $\tau_0\sim\ell_{11}^{-3}$.  
In fact, it will again be more convenient to calculate the correction 
$\tilde\tau=L\tau$ to the vacuum energy density integrated over the compact 
dimension, i.e., the effective string tension.  
The first contribution to $\tilde\tau$ comes again from the mismatch between 
the boundary conditions on bulk bosons and bulk fermions on the worldvolume, 
and does not involve the boundary $E_8$ current algebras.  Repeating the 
steps we used in our analysis of the spacetime Casimir effect in section~2.2, 
and taking into account that we have eight fermionic and eight bosonic 
degrees of freedom at each non-zero mass level, we obtain
\eqn\eeeffeten{\eqalign{\tilde\tau&=-\int\frac{d^2p}{(2\pi)^2}\sum_{p_i}
(-1)^{F_i}\int_0^\infty\frac{d\ell}{2\ell}e^{-(p^2+p_i^2)\ell/2}\cr
&\quad{}=-4\int_0^\infty\frac{d\ell}{2\ell}\frac{1}{2\pi\ell}
\theta_4(0|i\pi\ell/8L^2)\cr
&\qquad{}=-\frac{1}{L^2}\int_0^\infty\frac{dt}{8t^2}\theta_4(0|it).
\cr}}
This again has the expected Casimir form, and arguments similar to those in 
section~2.2 prove that the Casimir correction $\tilde\tau$ to the string 
tension, as given by \eeeffeten , is finite and negative.  This negative 
Casimir tension competes with the positive bare string tension $\tilde\tau_0
\sim L\ell_{11}^{-3}$.  While the supergravity approximation breaks down 
before we reach the regime of $L\sim\ell_{11}$, our results suggest that at 
distances $L$ smaller than the eleven-dimensional Planck scale, the effective 
string that corresponds to the stretched open membrane becomes tachyonic.  

\newsec{Applications}

Having demonstrated that the Casimir force between the boundaries of 
the $\EbarE$ model at large separations is attractive, we feel compelled to 
present a few remarks on the possible eventual fate of the $\EbarE$ 
configuration.  In the process, we will keep in mind two different 
perspectives:  the model can be viewed as an analog of the D-brane 
anti-D-brane systems of string theory, or alternatively, as a particular 
example of a brane-world compactification of M-theory with broken 
supersymmetry.  In this section, we offer a closer inspection of these two 
applications, before addressing the eventual fate of the $\EbarE$ system 
in section~4.  

\subsec{Analogy with the D$p$-D$\bar p$ systems}

The $\EbarE$ system in M-theory is in many ways analogous to the 
D$p$-D$\bar p$ brane systems recently studied in string theory.  Consider a 
system consisting of a certain number $N$ of coincident D$p$-branes separated 
by some distance $L$ from a system of $N$ coincident D$\bar p$-branes, for 
simplicity in flat $\R^{10}$.  This system differs from the BPS system of 
2$N$ D$p$-branes by the orientation reversal on the antibranes.  In this 
system, the branes and the antibranes each break a different half of the 
original supersymmetry, and the whole configuration is non-supersymmetric and 
unstable.  There is an attractive force between the branes and the antibranes 
\tomlenny , and at separations of order the string scale the $p\bar p$ open 
string connecting a D$p$-brane to a D$\bar p$-brane becomes tachyonic.  

All these facts have a close analogy in the $\EbarE$ system.  Indeed, the 
$\EbarE$ system differs from the BPS $E_8\times E_8$ system by the orientation 
reversal on the $\bar E_8$ boundary.  As we have demonstrated in section~2, 
there is an attractive Casimir force between the two boundaries.  
The closest M-theory analog of the $p\bar p$ open string stretching between 
D-branes is the open membrane stretching between the two boundaries.  The 
worldvolume Casimir effect found in section~2.4 suggests that the membrane 
becomes tachyonic at separations of order the Planck scale.  

This analogy becomes even more evident when we compactify one of the 
non-compact dimensions of the $\EbarE$ model on $\S^1$ with the supersymmetric 
spin structure, radius $R'$, and a Wilson line that breaks each $E_8$ to 
$SO(16)$, and then go to the limit of small $R'$ while keeping the distance 
$L$ between the boundaries large.  By cluster decomposition, this is 
equivalent to a $\Z_2$ orientifold of the weakly coupled Type IIA theory.  
This orientifold is a non-supersymmetric variant of the Type I${}'$ 
orientifold, with sixteen D8-branes on top of an orientifold plane at one end, 
and sixteen D$\bar 8$-branes on top of an orientifold plane with the opposite 
orientation (i.e., an ``antiorientifold'' plane) at the other end.  Clearly, 
the open stretched membrane connecting the $E_8$ and $\bar E_8$ boundaries 
descends to the open string stretched between the D8 and D$\bar 8$.  

In string theory, the D$p$-D$\bar p$ system is unstable, and is expected to 
decay to the supersymmetric vacuum \sen.  In the process, the open-string 
tachyon behaves as a Higgs field and condenses to a minimum of its potential, 
breaking the worldvolume gauge symmetry to its diagonal subgroup \ewk ,%
\eqn\eeubreak{U(N)\times U(N)\rightarrow U(N).}
Since the outcome of this annihilation should be equivalent to the 
supersymmetric vacuum, the residual gauge symmetry in \eeubreak\ should also 
disappear, presumably by the process suggested and analyzed in \sens .  This 
annihilation of the D$p$-D$\bar p$ system can be obstructed by the topological 
difference between the Chan-Paton bundles $E$ and $F$ carried by the 
D$p$-branes and D$\bar p$ branes.  The obstruction $E-F$ is naturally an 
element of the (reduced) K-theory group of spacetime, and can be interpreted 
as a lower-dimensional D-brane charge.  In this way, the spectrum of stable 
D-branes in codimension $k$ follows from the famous Bott periodicity pattern, 
$\tilde K(\S^k)=\Z$ for $k$ even, and zero for $k$ odd.  Alternatively, one 
can view the obstruction against annihilation as a topological defect in the 
tachyon field, classified by the homotopy groups of the vacuum manifold $U(N)$ 
of the worldvolume Higgs mechanism, 
\eqn\eehiggs{\eqalign{\pi_{2n-1}(U(N))&=\Z,\cr\pi_{2n}(U(N))&=0,\cr}}
with $N$ in the stable regime.  Using this representation, one can construct 
any stable D-brane (at least in the absence of the 3-form field strength, 
$H=0$) as a defect in the universal, spacetime-filling medium of sixteen 
D9-D$\bar 9$ pairs in Type IIB theory \ewk , or 32 unstable D9-branes 
in Type IIA theory \phk .%
\foot{For more details on the relation between D-brane charges, the 
worldvolume Higgs mechanism, and K-theory, see \refs{\ewk,\phk}.}

Once this picture is established in string theory, it is natural to ask 
whether it can be lifted to M-theory.  However, the spectrum of stable branes 
that one could use to build unstable brane systems in M-theory is very 
limited.  One could contemplate using M5-M$\bar 5$ pairs \yi , but such 
systems exhibit very complicated worldvolume dynamics, with Yang-Mills gauge 
bundles replaced by objects carrying two-form gauge fields.  In contrast, as 
we have just seen the $\EbarE$ system is a much closer M-theory analog of the 
D$p$-D$\bar p$ systems, in part also because the boundaries carry conventional 
Yang-Mills gauge bundles.  In fact, it turns out that the $\EbarE$ system 
exhibits certain properties expected of the universal system in M-theory.  

Since the two $E_8$ boundaries of the $\EbarE$ model attract each other one 
can imagine that in analogy with the D$p$-D$\bar p$ systems they could 
annihilate, possibly forming bound states whose conserved quantum numbers 
would be classified by the topological difference $E-F$ between the two 
$E_8$ bundles at the two boundaries.  Remarkably, $E_8$ bundles on a 
ten-manifold $M$ are classified by only one topological invariant 
$\lambda(E)\in H^4(M,\Z)$, which assigns an $E_8$ instanton number to each 
4-cycle in $M$.  Thus, the only topological difference between the two $E_8$ 
bundles $E$ and $F$ would be the difference between their ``instanton 
numbers,'' $\lambda(E)-\lambda(F)$.  Another way of seeing this follows from 
the structure of homotopy groups of the $E_8$ group manifold, which in the 
range of values of $k$ relevant for M-theory are given by \ewtools\ 
\eqn\eehomot{\eqalign{\pi_3(E_8)&=\Z,\cr\pi_k(E_8)&=0,\quad k\neq 3.\cr}} 
Even though we cannot follow the dynamics of the $\EbarE$ system to the 
regime of small separations $L\sim\ell_{11}$, the structure of the homotopy 
groups \eehomot\ and the analogy with the D$p$-D$\bar p$ systems suggest that 
at separations smaller than the Planck length the gauge symmetry should be 
broken to its diagonal subgroup, 
\eqn\eeeightbreak{E_8\times E_8\rightarrow E_8,}
Codimension $k$ defects in this Higgs pattern are topologically classified 
by the elements of the $(k-1)$-th homotopy group of the vacuum manifold 
$E_8$.   It is intriguing that \eehomot\ leaves precisely enough room for the 
M5-brane to appear as a bound state of two $E_8$ ends of the world!  Indeed, 
the quantum number in $\pi_3(E_8)$ can be interpreted as the M5-brane charge, 
since it corresponds to the difference between the $E_8$ instanton numbers at 
the two boundaries (on the four-cycle transverse to the defect).  This is in 
accord with the fact that a small $E_8$ instanton can leave the boundary in 
M-theory as a bulk M5-brane.  

We conclude our discussion of the analogy with D$p$-D$\bar p$ systems with a 
few remarks: 

\item{(1)} Since the $\EbarE$ system of M-theory is so closely related to 
D$p$-D$\bar p$ systems of string theory, it is natural to expect that as the 
two $E_8$ boundaries come close together under the influence of the Casimir 
force, some form of brane-antibrane annihilation will take place.  We will 
present further evidence supporting this conjecture in section~4.  

\item{(2)} Further compactification on $\S^1$ with the supersymmetric spin 
structure allows one to interpret the $\EbarE$ system as a natural lift to 
M-theory of the system of sixteen D8-D$\bar 8$ pairs.  Note that this is 
precisely the most natural value suggested by K-theory for the universal 
system of unstable branes in Type IIA string theory \refs{\ewk,\phk}, and 
the $\EbarE$ system is large enough to be universal in M-theory.  

\item{(3)} If the gauge symmetry is broken at small $L$ according to 
\eeeightbreak , a residual $E_8$ gauge symmetry survives.  In the low-energy 
field theory approximation of the D$p$-D$\bar p$ system, the Higgs pattern 
\eeubreak\ leaves a residual non-supersymmetric $U(N)$ Yang-Mills theory on 
top of the supersymmetic vacuum, whose fate in the full theory is discussed 
in \sens .  Is there a candidate for describing the residual $E_8$ gauge 
symmetry in M-theory?  Since the size of the eleventh dimension is small 
in the Higgs regime, such a description -- if it exists -- should be in terms 
of a weakly coupled, non-supersymmetric, non-chiral, and modular invariant 
heterotic string theory in $\R^{10}$ with gauge group $E_8$.  It is intriguing 
that a heterotic string theory with such properties does in fact exist 
\refs{\klt,\dixh}.  Its perturbative spectrum contains no tachyons in 
non-trivial $E_8$ representations, but there is a neutral tachyon suggesting 
a residual instability of this theory, which could be related to the inherent 
instability of the $\EbarE$ system discussed in section~4.  

\subsec{Brane-world scenarios}

Compactify the $\EbarE$ model down to $\R^4\times\S^1/\Z_2$ on some 
six-manifold $Y$.  One of the boundaries can then be viewed as a 
``brane-world,'' and the whole system as a model for supersymmetry breaking 
in the brane-world scenario.  Similar compactifications of the 
supersymmetric $E_8\times E_8$ theory on $Y$ which is (to zeroth-order) a 
Calabi-Yau manifold preserve $\CN=1$ supersymmetry in the four non-compact 
dimensions \ewcy .   One can similarly compactify the $\EbarE$ model on such 
$Y$ so that each boundary preserves $\CN=1$ supersymmetry.  However, due to 
the mismatch between the two boundaries, all supersymmetries are broken 
in the full system.  This pattern of supersymmetry breaking is very similar 
to the supersymmetry breaking in the supersymmetric $E_8\times E_8$ theory by 
gluino condensation in the hidden $E_8$ \phg:  in that case, the gluino 
condensate at the hidden boundary still preserves $\CN=1$ supersymmetry, which 
is however mismatched with the $\CN=1$ supersymmetry preserved at the other 
boundary.  

Previous studies of supersymmetry breaking patterns in heterotic M-theory 
(such as the hidden sector supersymmetry breaking of \phg) lead us to expect 
the M-theoretic dual of the dilaton runaway problem \refs{\runaway,\bdstrong} 
-- for large initial distances $L$ between the boundaries, the potential for 
$L$ tends to run $L$ to infinity, and therefore zero effective coupling 
$1/L$.  In contrast, the Casimir effect in the $\EbarE$ model drives 
$L$ to smaller values, and can therefore play an important role in the 
radius stabilization problem.  This issue clearly deserves a closer study  
of the Casimir effect in compactifications of the $\EbarE$ model on $Y$, which 
is beyond the scope of the present paper.  

\newsec{Fate of the $\EbarE$ System: End-of-the-World Annihilation in M-Theory}

As we have seen, the $\EbarE$ system is a close analog of the unstable 
D$p$-D$\bar p$ systems of string theory, and one may expect that the eventual 
fate of the system will involve some form of brane-antibrane annihilation.    
Upon further compactification on an extra $\S^1$ with the supersymmetric spin 
structure, the two $E_8$ boundaries indeed descend to a system of sixteen 
D8-branes and sixteen D$\bar 8$-branes on top of two orientifold 8-planes, and 
we certainly expect the D8-D$\bar 8$ system to annihilate.  When lifted to 
M-theory, this expectation immediately leads to a puzzle: assuming that the 
$E_8$ degrees of freedom at the two boundaries annihilate, what is left after 
this annihilation?  Are we left with some M-theory analogs of orientifold 
planes with no Yang-Mills degrees of freedom?  Such orientifold planes would 
carry a gravitational anomaly \hw .  Or do the orientifold planes also 
annihilate each other in the process, restoring M-theory on $\S^1$ with some 
(small) radius and 32 supersymmetries?  

These questions are of course difficult to address directly because the 
answers lie in the strongly coupled regime where we have no control over the 
theory.%
\foot{It does not seem possible to use a matrix model definition of the 
$\EbarE$ system, due to the difficulty one would have with defining a 
light-cone frame in the metric that is curved by the Casimir effect, and due 
to the absence of supersymmetry needed to protect flat directions and hence 
a macroscopic spacetime in matrix theory.}
It turns out, however, that we can study the fate of the system already at 
large $L$, where the annihilation of the two boundaries is a non-perturbative 
effect suppressed exponentially in (a power of) $1/L$.  As we are now going to 
show, this argument reveals that neither of the two scenarios outlined above 
are realized.  It turns out that the $\EbarE$ system is unstable to false 
vacuum decay \hhkfalse , which is of the catastrophic type \ewkk\ with 
the spacetime manifold annihilating to nothing!  

\subsec{The wormhole instanton}

Consider the $\EbarE$ model on $\R^{10}\times\S^1/\Z_2$ with the flat, direct 
product metric%
\eqn\eeflatsln{ds^2_0=\eta_{MN}dx^Mdx^N,\qquad x^{10}\in[0,L].}
This configuration represents a classical solution, whose first quantum 
corrections in the long-wavelength, large-$L$ expansion due to the Casimir 
effect were calculated in section~2.3.  There is a Euclidean instanton in this 
theory, asymptotic to \eeflatsln\ as $r\equiv\sqrt{\eta_{AB}x^Ax^B}\rightarrow
\infty$.  This instanton is given by a $\Z_2$ orbifold of the Euclidean 
Schwarzschild solution in eleven dimensions,%
\foot{All of our gravity sign conventions are as in Misner, Thorne and 
Wheeler, \mtw.}
\eqn\eeeucsch{ds^2=\left(1-\left(\frac{4L}{\pi r}\right)^8\right)
(dx^{10})^2+\frac{dr^2}{1-\left(\frac{4L}{\pi r}\right)^8}+r^2d^2\Omega_9,}
under the orbifold action $x^{10}\rightarrow -x^{10}$.  
The Euclidean Schwarzschild solution indeed has the correct spin structure 
asymptotically at large $r$ (recall \eeantipgrino ), and also survives 
the $\Z_2$ projection; hence, it represents a legitimate classical 
solution of the $\EbarE$ compactification of M-theory asymptotic to 
\eeflatsln , in the supergravity approximation.%
\foot{In string theory, similar orbifolds of Euclidean Schwarzschild black 
holes (in $1+1$ dimensions) were considered in \phbh .  In that case, $\Z_2$ 
is an orientifold symmetry, which reverses worldsheet orientation.  Similarly 
in the present case, if we compactify \eeeucsch\ on an extra $\S^1$ with the 
supersymmetric spin structure, it corresponds to an orientifold of the 
Schwarzschild solution of Type IIA theory.}
While the Euclidean Schwarzschild solution is topologically $\R^2\times\S^9$, 
its $\Z_2$ orbifold is topologically $\R^2_+\times\S^9$, where $\R^2_+$ 
denotes the half-plane.  Thus, this solution has only one boundary component, 
topologically $\R\times\S^9$.  

The Euclidean Schwarzschild instanton has a negative mode, which also 
survives our orbifold projection.  Since \eeeucsch\ is smooth and falls off 
fast enough at inifinity to have zero ADM mass, the positive energy theorem 
is manifestly invalid classically in the $\EbarE$ system.  The 
instanton \eeeucsch\ represents a bounce, responsible for false vacuum decay 
in the theory.  (For some background on false vacuum decay in field theory,  
gravity, and string theory, see \refs{\coleman,\ewkk,\hhk} and \hhkfalse .)  

The outcome of the false vacuum decay mediated by the bounce instanton 
\eeeucsch\ can be read off from the turning point of the instanton and its 
subsequent evolution in the Minkowski signature.  The turning point of 
\eeeucsch\ can be identified as follows.  Write the metric on the $\S^9$ as 
$d^2\Omega_9=d\theta^2+\sin^2\theta\,d^2\Omega_8$, where $d^2\Omega_8$ is the 
round metric on $\S^8$, and $\theta\in[0,\pi]$.  The turning point 
corresponds to $\theta=\pi/2$, a slice of space with zero extrinsic curvature, 
topologically $\R_+\times\S^8$.  Thus, the geometry nucleated by the 
instanton \eeeucsch\ has the form of a wormhole connecting the two 
boundaries, as depicted in Fig.~3.  The evolution of this initial condition 
is obtained by Wick-rotating the Euclidean time $\theta\rightarrow\pi/2+it$.  
At $t>0$, the Minkowski-signature metric is
\eqn\eeblkmink{ds^2=-r^2dt^2+\left(1-\left(\frac{4L}{\pi r}\right)^8\right)
(dx^{10})^2+\frac{dr^2}{1-\left(\frac{4L}{\pi r}\right)^8}+r^2\cosh^2t\,d^2
\Omega_8.}
It is convenient to introduce new coordinates $(W,T)$, given by 
\eqn\eecoordnew{\eqalign{W&=r\cosh t,\cr T&=r\sinh t.\cr}}
In these coordinates, the metric on the boundary $x^{10}=0$ of \eeblkmink\ 
becomes 
\eqn\eeboundmink{ds^2=-dT^2+dW^2+\left(\pi^8\left(\frac{W^2-T^2}{16L^2}
\right)^4-1\right)^{-1}\frac{(WdW-TdT)^2}{W^2-T^2}+W^2d^2
\Omega_8.}
Our coordinate system is singular at $W^2=16L^2/\pi^2+T^2$ and describes only 
one half of the full, smooth geometry of the expanding wormhole.  The other 
half of the wormhole is a mirror copy of \eeboundmink , and connects smoothly 
to \eeboundmink\ at the eight-sphere $\S^8_{min}$ of minimal area located 
at $W^2=16L^2/\pi^2+T^2$ inside the wormhole.  The radius $\CR_{min}$ of the 
minimal-area sphere $\S^8_{min}$ increases with growing $T$, with a speed 
approaching the speed of light: 
\eqn\eegrowth{\CR_{min}(T)=\sqrt{16L^2/\pi^2+T^2}.}
\fig{The wormhole geometry nucleated at $t=0$ by the ``bounce'' \eeeucsch .  
After its nucleation, the size of the wormhole expands with a speed 
quickly approaching the speed of light.}{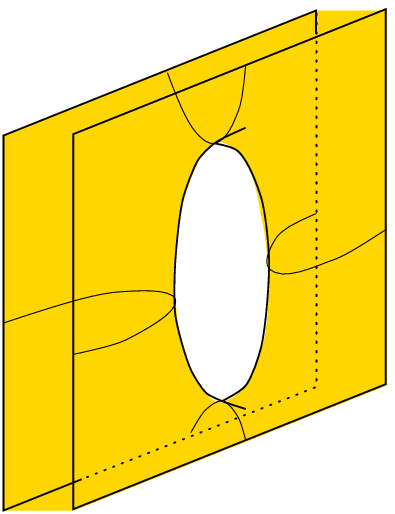}{2truein}

The existence of the wormhole bounce solution \eeeucsch\ in the $\EbarE$ 
model indicates the existence of a decay channel in which the vacuum decays to 
nothing by nucleating wormholes.  The probability for nucleating a single 
wormhole per unit boundary area and unit time is exponentially small in $1/L$, 
and of order 
\eqn\eeexpon{\exp\left\{-\frac{4(2L)^8}{3\pi^4 G_{10}}\right\}}
where $G_{10}$ is the ten-dimensional effective Newton constant.  The exponent 
in \eeexpon\ is one half of the action of the Euclidean Schwarzschild black 
hole in eleven dimensions, since our bounce corresponds to one half of the 
full black hole geometry.  

Thus, we have discovered a non-perturbative mechanism which indeed corresponds 
to the expected annihilation between the $E_8$ and $\bar E_8$ boundaries.   
This also resolves the small puzzle raised at the beginning of this 
section: in this annihilation process, not only the $E_8$ ``branes'' 
annihilate -- the whole spacetime does!  

This spacetime annihilation is a non-perturbative effect in $1/L$.  So far 
in this section, we have neglected perturbative corrections in powers of 
$1/L$.  Indeed, the perturbative Casimir effect will dominate at large $L$ 
over the exponentially suppressed decay of the (approximate) vacuum.  If 
the full potential has a minimum at some large value of $L$, the bounce 
solution will be slightly modified, but we expect our conclusions about the 
catastrophic instability of this vacuum to hold.  If the dynamics of the 
system drives $L$ to values of order the Planck scale, our approximation  
becomes invalid.  However, if the system settles in a minimum of the potential 
outside the reach of large-$L$ perturbation theory without encountering a 
phase transition, this minimum will still be separated from the catastrophic 
decay to nothing by only a finite-size potential barrier.  

\subsec{Boundary geometry of the expanding wormhole}

Once a wormhole is nucleated, it will expand with a speed approaching the 
speed of light, at least until multi-wormhole effects become relevant.  We 
will now look more closely at the geometry induced on the boundary of a single 
expanding wormhole \eeblkmink .  

The bulk geometry \eeblkmink\ describes a non-static metric which satisfies 
the vacuum Einstein equations $R_{MN}=0$.  On the other hand, the metric 
induced on the boundary is not Ricci flat; a straightforward calculation 
reveals 
\eqn\eericcibound{\tilde R_{AB}=4\left(\frac{4L}{\pi}\right)^8
\frac{1}{r^{10}}\left(g_{AB}-10\,\hat e^r_A\hat e^r_B\right),}
where $\hat e^r$ is the unit one-form along $dr$, 
\eqn\eeunite{\hat e^r_A=\frac{\delta^r_A}{\sqrt{1-\left(\frac{4L}{\pi r}
\right)^8}},}
and $\tilde R_{AB}$ is the Ricci tensor of the boundary metric (not to be 
confused with the $AB$ components of the bulk Ricci tensor $R_{MN}$.) 
Notice that the coefficient in front of the $\hat e^r\hat e^r$ term in 
\eericcibound\ is precisely such that the boundary Ricci scalar vanishes, 
\eqn\eericcizero{\tilde R\equiv g^{AB}\tilde R_{AB}=0.}
Thus, the boundary observer perceives an expanding universe, and feels the 
presence of an effective matter distribution whose energy-momentum tensor 
is traceless, 
\eqn\eetabeffect{T_{AB}=\frac{1}{8\pi\tilde G_{10}}\tilde R_{AB}=
\frac{1}{2\pi\tilde G_{10}}\left(\frac{4L}{\pi}\right)^8\frac{1}{r^{10}}
\left(g_{AB}-10\,\hat e^r_A\hat e^r_B\right),}
with $\tilde G_{10}$ the effective Newton constant at the boundary.  Notice 
the characteristic non-perturbative behavior $T_{AB}\sim L^8/\tilde G_{10}$.  

The boundary Ricci scalar $\tilde R$ vanishes, but there will be other 
curvature invariants that are non-zero.  For example, one finds 
\eqn\eenonzero{\tilde R_{AB}\tilde R^{AB}=1440\left(\frac{4L}{\pi}\right)^{16}
\frac{1}{r^{20}}.}
This curvature invariant reaches its largest value on the smallest sphere 
inside the wormhole at the nucleation time $T=0$, where it is equal to 
$\frac{45\pi^4}{8L^4}$.  This is indeed small for large $L$ and our 
supergravity approximation is valid everywhere as long as the asymptotic 
separation of the two boundaries is large.  

One can introduce a coordinate $y$ that is better suited to study the geometry 
of the expanding wormhole near its center $\S^8_{min}$, 
\eqn\eexcoord{y=\sqrt{1-\left(\frac{4L}{\pi r}\right)^8}.}
This coordinate covers the inside of the wormhole, with the sphere 
$\S^8_{min}$ of minimal area at $y=0$.  One can now express the boundary 
metric near $y\approx 0$ as follows,
\eqn\eewrmmet{\eqalign{ds^2&\approx\left(\frac{4L}{\pi}\right)^2\left\{
\left(1-\frac{1}{4}y^2+\ldots\right)(-dt^2+\cosh^2t\,d^2\Omega_8)\right.\cr
&\left.\qquad\qquad{}+\frac{1}{16}\left(1-\frac{9}{4}y^2+\ldots\right)dy^2
\right\}.\cr}}
Thus, at large proper times since the nucleation of the wormhole, the observer 
located inside the wormhole at $y=0$ will experience exponential inflation of 
the wormhole throat $\S^8$.  

It is also instructive to calculate $\delta(x^{10})\wedge\tr\,(R\wedge R)$, 
since this expression appears in the Bianchi identity \hweff\ for the 
four-form field strength $G$ and, if non-zero, serves as a source for $G$.  
However, it is straightforward to see that the four-form $\omega\equiv\tr\,
(R\wedge R)$ at the boundary is zero.  Indeed, $\omega$ can be written as a 
sum $\omega=\sum_{p=1}^4\omega_p$, where $\omega_p$ is a four-form with $p$ 
of its legs on $\S^8$ (and possibly dependent on the coordinates transverse 
to the $\S^8$); but no such invariant $p$-forms exist on $\S^8$ with the round 
metric, and $\delta(x^{10})\wedge\tr\,(R\wedge R)$ vanishes for the boundary 
geometry given by \eeboundmink .  

\newsec{Conclusions}

In this paper we have demonstrated the existence of an attractive Casimir 
force between two $E_8$ boundaries with mismatched chiralities in M-theory.  
In fact, we have argued that -- in analogy with the D$p$-D$\bar p$ brane 
systems of string theory -- the two boundaries of the $\EbarE$ system 
annihilate, in a process which annihilates the entire spacetime manifold to 
nothing.  

From the point of view of the bulk observer, this is just another example of 
the catastrophic false vacuum decay \refs{\ewkk,\hhkfalse} whereby a hole in 
the spacetime manifold is first nucleated and then expands with a speed 
approaching the speed of light.  As a consequence, we would not want to live 
in the bulk.  

For a boundary observer, however, the decay of the $\EbarE$ system looks 
a little less catastrophic:  a wormhole connecting the two 
boundaries is nucleated, and the radius of its throat expands exponentially.  
Thus, living inside the boundary is perhaps not as bad as living in the bulk.  
The boundary observer indeed experiences the decay of the bulk as a 
time-dependent cosmological evolution of the boundary, and observes 
topology-changing processes that connect the observed brane-world to its 
hidden counterpart.  This is an example of what one should expect in general 
``many-fold'' universe scenarios such as those of \manyf , where the 
neighboring folds of the brane-world are each other's antibranes.  

This instability to false vacuum decay is rather generic in non-supersymmetric 
compactifications of string theory and M-theory \hhkfalse , and could impose 
a strong constraint on phenomenologically acceptable scenarios.  In the case 
of brane-worlds, one could prevent catastrophic vacuum decay by considering 
non-supersymmetric branes that carry a K-theory charge.  It is perhaps 
not necessary to look for a compactification where the catastrophic decay is 
absent, however.  Indeed, in the $\EbarE$ model at large boundary separations 
$L$, the wormhole nucleation -- and therefore the probability for spacetime 
to decay into nothing -- is exponentially suppressed with $1/L$ (see 
\eeexpon ), and for large enough $L$, the lifetime of the universe can still 
be cosmologically large.  This creates an intriguing possibility whereby the 
cosmological evolution of the observed universe would correspond to the 
evolution on the boundary of a bulk spacetime undergoing a catastrophic 
vacuum decay!  

In analogy with the D$p$-D$\bar p$ systems of string theory, we expect the 
two $E_8$ boundaries of the $\EbarE$ system to completely annihilate only if 
there is no topological obstruction carried by the two $E_8$ bundles.  In 
section~3 we presented topological arguments suggesting that the system can 
support 5-brane bound states.  It is tempting to speculate that the bulk 
spacetime of the $\EbarE$ system with a non-zero net 5-brane charge would 
still annihilate, possibly leaving behind the 5-brane charges in the form of a 
little string theory.  

\bigskip\medskip
During the course of this work, M.F.~benefitted from helpful discussions with 
J.~Form\'anek, L.~Motl, and J.~Niederle.  P.H.~wishes to acknowledge 
useful conversations with O.~Bergman, E.~Gimon, C.~Johnson, and E.~Witten.  
The work of M.F. has been supported in part by an undergraduate fellowship at 
the Faculty of Mathematics and Physics, Charles University, and by the 
Institute of Physics, Academy of Sciences of the Czech Republic under Grant 
No.\ GA-AV\v CR A1010711.  The work of P.H. has been supported by a Sherman 
Fairchild Prize Fellowship, and by DOE grant DE-FG03-92-ER~40701.  
\listrefs
\end